\documentclass[a4paper,fleqn]{article}
\usepackage{INTERSPEECH_v2,amssymb,amsmath,epsfig}
\usepackage{xspace}
\usepackage{srcltx}
\usepackage{caption}
\usepackage{subcaption}
\usepackage[export]{adjustbox}
\usepackage{multirow}
\usepackage{tikz}
\usepackage{stfloats}

\ninept

\def\reg{{\rm\ooalign{\hfil
     \raise.07ex\hbox{\scriptsize R}\hfil\crcr\mathhexbox20D}}}

\definecolor{darkspringgreen}{rgb}{0.09, 0.55, 0.27}

\usepackage{color}

\usepackage[final]{changes}

\def\vx{\mathbf{x}}

\def\vy{\mathbf{y}}

\def\ve{\mathbf{e}}

\def\vc{\mathbf{c}}

\def\reg{{\rm\ooalign{\hfil
     \raise.07ex\hbox{\scriptsize R}\hfil\crcr\mathhexbox20D}}}

\def\vx{\mathbf{x}}

\ninept
\begin{document}
\title{A Deliberation-based Joint Acoustic and Text Decoder}
\name{Sepand Mavandadi, Tara N. Sainath, Ke Hu, Zelin Wu}
\address{Google Inc., U.S.A}
\email{\{sepand, tsainath, huk, zelinwu\}@google.com}

\maketitle
\begin{abstract}
We propose a new two-pass E2E speech recognition model that improves ASR performance by training on a combination of paired data and unpaired text data. Previously, the joint acoustic and text decoder (JATD) has shown promising results through the use of text data during model training and the recently introduced deliberation architecture has reduced recognition errors by leveraging first-pass decoding results. Our method, dubbed Deliberation-JATD, combines the spelling correcting abilities of deliberation with JATD's use of unpaired text data to further improve performance. The proposed model produces substantial gains across multiple test sets, especially those focused on rare words, where it reduces word error rate (WER) by between 12\% and 22.5\% relative. This is done without increasing model size or requiring multi-stage training, making Deliberation-JATD an efficient candidate for on-device applications.

\end{abstract}

\section{Introduction \label{sec:introduction}}

E2E models \cite{Ryan19,CC18,Graves12,GravesMohamedHinton13,RaoSakPrabhavalkar17,Chan15,kim2017ctcattn,ChiuRaffel17} combine the acoustic (AM), pronunciation (PM) and language models (LM) of a conventional ASR system into a single neural network. This structure makes them significantly smaller than conventional models \cite{CC18,SainathPang19} and ideal for on-device ASR \cite{Ryan19}. However, the performance of E2E models on rare word recognition still lags behind conventional models. The performance gap is partially because they lack the ability to train using text-only data, which is abundant and often utilized by conventional LMs.

There have been multiple approaches for augmenting E2E models and training procedures to incorporate unpaired text data. Broadly speaking, these approaches use some combination of an LM trained on text data (shallow, cold, deep fusion \cite{Jan17,Sriram17,Anjuli18,inaguma2019fusion}) and a multi-stage training procedure that incorporates unpaired data (``weak distillation" \cite{Bo19}, ``back-translation" \cite{Sennrich16}, ``cycle-consistency" \cite{Hori19,DBLP:conf/icml/RenTQZZL19, Bai2019}). Each approach produces improvements in performance, but also increases some combination of model size, training and inference complexity, making it less desirable for on-device applications.

A recent approach, the joint acoustic and text decoder (JATD) \cite{sainath2020jatd, wang2020jatd} side-steps these issues. It utilizes unpaired data during E2E model training either directly or by generating the missing half of the data; using TTS to generate audio from text and ASR to generate text from audio. Previous methods using unpaired data in training have shown limited success in improving performance on real audio test sets \cite{Bo19,Ding19}. JATD produces stronger results by using a fixed context vector as a ``domain ID" to distinguish between paired and unpaired data during training. Paired data is processed as normal with the encoder computing an acoustic context vector that is fed to the decoder. For unpaired data (text-only or with synthesized audio), JATD bypasses the encoder network by using a fixed but learnable context vector in place of the encoder output, allowing the model to train on text-only data and avoiding the encoder training on synthesized audio.

The JATD architecture results in only a trivial increase in model size. It has been implemented within a LAS two-pass decoding framework and trained on both audio-text pair data and unpaired text-only data (by synthesizing into TTS utterances), showing substantial improvements in WER, especially on rare words \cite{sainath2020jatd}.

Beside data augmentation approaches, novel model architectures also show improvements in recognizing rare words. Recently, deliberation models \cite{xia2017deliberation, hu2020deliberation, hu2021deliberation} have used an attention-based two-pass design to achieve state-of-the-art performance on Google VoiceSearch test sets. Similar to other two-pass models, deliberation uses its first-pass decoder to produce streaming hypotheses and its second-pass decoder to attend to the first-pass hypotheses alongside the encoder outputs for redecoding or rescoring. The hypothesis attention allows deliberation to act like a spelling corrector on full-context first-pass hypotheses. This results in substantial performance gains, especially on rare words \cite{hu2020deliberation, hu2021deliberation}.

Our novel contribution is to combine the text-only training capabilities of JATD with the spelling-correction benefits of deliberation. Our approach, dubbed deliberation-JATD, augments deliberation's attention contexts to use JATD's fixed context vectors, enabling the architecture to train on text-only data. Experiments show deliberation-JATD improving rare word performance by at least 12\% relative to both LAS-JATD and deliberation without any degradation on VoiceSearch tasks.

\section{Methods} \label{sec:model}

In this section, we describe our baseline approaches, deliberation and LAS-JATD, as well as the proposed Deliberation-JATD method.

\subsection{Baseline Approaches}

\subsubsection{Deliberation}
\label{sec:methods_delib}

\begin{figure}[h!]
\begin{center}

\centering \scalebox{0.8}{
\input{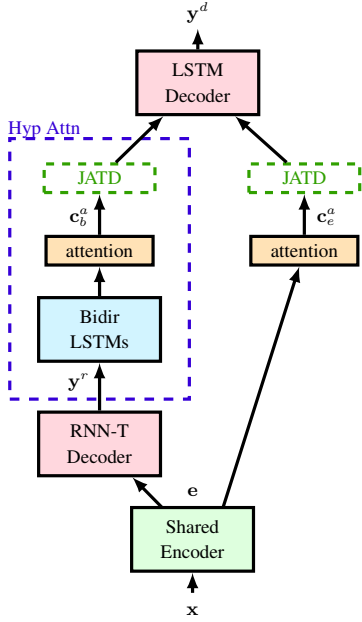}
}
\end{center}
\caption[]{\label{fig:delib} 2-pass model architecture. With ``Hyp Attn'' block: a deliberation model. Without "Hyp Attn" block: a 2-pass LAS model. Including the JATD block only on the right branch results in the ``partial'' variant of Deliberation-JATD. Including both JATD blocks results in the ``full'' variant.
}
\end{figure}

Deliberation networks (fig. \ref{fig:delib}) consist of a shared encoder, a first-pass RNN-T decoder, and a second-pass deliberation decoder. The shared encoder takes log-mel filterbank energies, $\vx=(\vx_1 \ldots \vx_T)$, where $T$ denotes the number of frames, and generates an encoding, $\ve$. This encoder output, $\ve$, is then fed to an RNN-T decoder to produce first-pass decoding results, $\vy^r$, in a streaming fashion. The deliberation decoder attends to both $\ve$ and $\vy^r$, producing two context vectors, $\vc_e^a$ and $\vc_b^a$, that are concatenated and passed as inputs to an attention-based LSTM decoder. This decoder produces the final probabilities, $\vy^d$, which can be written as $p(\vy^d | \vx, \vc_e^a, \vc_b^a , \vy_{u - 1:1}^d )$, where $\vy_{u - 1:1}^d = {\vy_{u - 1}^d , \ldots , \vy_1^d }$ indicates all previous decoded labels of a single hypothesis during inference.

Inference for deliberation models is done in two passes. First, the RNN-T decoder processes encoder outputs, $\ve$, to produce the first-pass sequence, $\vy^r$. Then, the deliberation decoder attends to $\ve$ and the complete first pass hypotheses, $\vy^r$, and performs a second beam search to generate $\vy^d$. This second pass acts as a spell-corrector, using the full context of the first pass hypothesis to substantially improve performance \cite{hu2020deliberation}.

Deliberation training requires audio-text pairs and does not offer a natural way to incorporate unpaired text data. In the following section, we describe JATD, which addresses this shortfall.
\subsubsection{LAS-JATD
\label{sec:methods_las_jatd}}

JATD was implemented \cite{sainath2020jatd} on a two-pass LAS model running beam search in the second-pass decoder. This two-pass LAS model can be succinctly described as a deliberation model where the second-pass LSTM decoder does not use the the first-pass RNN-T decoder outputs, $\vy^r$, and only attends to the shared encoder outputs, $\ve$. This is equivalent to fig. \ref{fig:delib} with the ``Hyp Attn'' block removed. The final log probabilities output by LAS can be written as $\log{p(\vy_u^d |\vx^a, \vc_e^a , \vy_{u - 1:1}^d )}$.

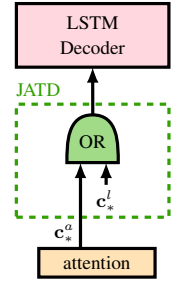
\begin{figure}[h!]
\begin{center}
\centering \scalebox{0.8}{
\ifx\du\undefined
  \newlength{\du}
\fi
\setlength{\du}{15\unitlength}
\begin{tikzpicture}[even odd rule]
\pgftransformxscale{1.000000}
\pgftransformyscale{-1.000000}
\definecolor{dialinecolor}{rgb}{0.000000, 0.000000, 0.000000}
\pgfsetstrokecolor{dialinecolor}
\pgfsetstrokeopacity{1.000000}
\definecolor{diafillcolor}{rgb}{1.000000, 1.000000, 1.000000}
\pgfsetfillcolor{diafillcolor}
\pgfsetfillopacity{1.000000}
\pgfsetlinewidth{0.100000\du}
\pgfsetdash{{0.200000\du}{0.200000\du}}{0\du}
\pgfsetmiterjoin
{\pgfsetcornersarced{\pgfpoint{0.000000\du}{0.000000\du}}\definecolor{diafillcolor}{rgb}{0.000000, 0.000000, 0.000000}
\pgfsetfillcolor{diafillcolor}
\pgfsetfillopacity{0.000000}
\fill (4.200000\du,3.763660\du)--(4.200000\du,7.317587\du)--(8.958597\du,7.317587\du)--(8.958597\du,3.763660\du)--cycle;
}{\pgfsetcornersarced{\pgfpoint{0.000000\du}{0.000000\du}}\definecolor{dialinecolor}{rgb}{0.192157, 0.627451, 0.000000}
\pgfsetstrokecolor{dialinecolor}
\pgfsetstrokeopacity{1.000000}
\draw (4.200000\du,3.763660\du)--(4.200000\du,7.317587\du)--(8.958597\du,7.317587\du)--(8.958597\du,3.763660\du)--cycle;
}% setfont left to latex
\definecolor{dialinecolor}{rgb}{0.192157, 0.627451, 0.000000}
\pgfsetstrokecolor{dialinecolor}
\pgfsetstrokeopacity{1.000000}
\definecolor{diafillcolor}{rgb}{0.192157, 0.627451, 0.000000}
\pgfsetfillcolor{diafillcolor}
\pgfsetfillopacity{1.000000}
\node[anchor=base east,inner sep=0pt, outer sep=0pt,color=dialinecolor] at (8.608597\du,5.734686\du){};
\pgfsetlinewidth{0.100000\du}
\pgfsetdash{}{0pt}
\pgfsetmiterjoin
\pgfsetbuttcap
\definecolor{diafillcolor}{rgb}{0.631373, 0.858824, 0.533333}
\pgfsetfillcolor{diafillcolor}
\pgfsetfillopacity{1.000000}
\definecolor{dialinecolor}{rgb}{0.000000, 0.000000, 0.000000}
\pgfsetstrokecolor{dialinecolor}
\pgfsetstrokeopacity{1.000000}
\pgfpathmoveto{\pgfpoint{7.377110\du}{5.600000\du}}
\pgfpathcurveto{\pgfpoint{6.843780\du}{5.600000\du}}{\pgfpoint{6.310440\du}{5.600000\du}}{\pgfpoint{5.777110\du}{5.600000\du}}
\pgfpathcurveto{\pgfpoint{5.777110\du}{5.000000\du}}{\pgfpoint{5.777110\du}{4.100000\du}}{\pgfpoint{6.577110\du}{4.100000\du}}
\pgfpathcurveto{\pgfpoint{7.377110\du}{4.100000\du}}{\pgfpoint{7.377110\du}{5.000000\du}}{\pgfpoint{7.377110\du}{5.600000\du}}
\pgfpathclose
\pgfusepath{fill,stroke}
\pgfsetlinewidth{0.100000\du}
\pgfsetdash{}{0pt}
\pgfsetmiterjoin
{\pgfsetcornersarced{\pgfpoint{0.000000\du}{0.000000\du}}\definecolor{diafillcolor}{rgb}{1.000000, 0.843137, 0.870588}
\pgfsetfillcolor{diafillcolor}
\pgfsetfillopacity{1.000000}
\fill (4.201490\du,0.600000\du)--(4.201490\du,2.600000\du)--(8.957108\du,2.600000\du)--(8.957108\du,0.600000\du)--cycle;
}{\pgfsetcornersarced{\pgfpoint{0.000000\du}{0.000000\du}}\definecolor{dialinecolor}{rgb}{0.000000, 0.000000, 0.000000}
\pgfsetstrokecolor{dialinecolor}
\pgfsetstrokeopacity{1.000000}
\draw (4.201490\du,0.600000\du)--(4.201490\du,2.600000\du)--(8.957108\du,2.600000\du)--(8.957108\du,0.600000\du)--cycle;
}% setfont left to latex
\definecolor{dialinecolor}{rgb}{0.000000, 0.000000, 0.000000}
\pgfsetstrokecolor{dialinecolor}
\pgfsetstrokeopacity{1.000000}
\definecolor{diafillcolor}{rgb}{0.000000, 0.000000, 0.000000}
\pgfsetfillcolor{diafillcolor}
\pgfsetfillopacity{1.000000}
\node[anchor=base,inner sep=0pt, outer sep=0pt,color=dialinecolor] at (6.579299\du,1.394062\du){LSTM};
% setfont left to latex
\definecolor{dialinecolor}{rgb}{0.000000, 0.000000, 0.000000}
\pgfsetstrokecolor{dialinecolor}
\pgfsetstrokeopacity{1.000000}
\definecolor{diafillcolor}{rgb}{0.000000, 0.000000, 0.000000}
\pgfsetfillcolor{diafillcolor}
\pgfsetfillopacity{1.000000}
\node[anchor=base,inner sep=0pt, outer sep=0pt,color=dialinecolor] at (6.579299\du,2.194063\du){Decoder};
\pgfsetlinewidth{0.100000\du}
\pgfsetdash{}{0pt}
\pgfsetbuttcap
{
\definecolor{diafillcolor}{rgb}{0.000000, 0.000000, 0.000000}
\pgfsetfillcolor{diafillcolor}
\pgfsetfillopacity{1.000000}

\pgfsetarrowsend{latex}
\definecolor{dialinecolor}{rgb}{0.000000, 0.000000, 0.000000}
\pgfsetstrokecolor{dialinecolor}
\pgfsetstrokeopacity{1.000000}
\draw (6.583080\du,4.096640\du)--(6.579300\du,2.600000\du);
}
% setfont left to latex
\definecolor{dialinecolor}{rgb}{0.192157, 0.627451, 0.000000}
\pgfsetstrokecolor{dialinecolor}
\pgfsetstrokeopacity{1.000000}
\definecolor{diafillcolor}{rgb}{0.192157, 0.627451, 0.000000}
\pgfsetfillcolor{diafillcolor}
\pgfsetfillopacity{1.000000}
\node[anchor=base west,inner sep=0pt,outer sep=0pt,color=dialinecolor] at (4.200000\du,3.500000\du){JATD};
\pgfsetlinewidth{0.100000\du}
\pgfsetdash{}{0pt}
\pgfsetbuttcap
{
\definecolor{diafillcolor}{rgb}{0.000000, 0.000000, 0.000000}
\pgfsetfillcolor{diafillcolor}
\pgfsetfillopacity{1.000000}

\pgfsetarrowsend{latex}
\definecolor{dialinecolor}{rgb}{0.000000, 0.000000, 0.000000}
\pgfsetstrokecolor{dialinecolor}
\pgfsetstrokeopacity{1.000000}
\draw (6.200000\du,8.300000\du)--(6.200000\du,5.600000\du);
}
\pgfsetlinewidth{0.100000\du}
\pgfsetdash{}{0pt}
\pgfsetbuttcap
{
\definecolor{diafillcolor}{rgb}{0.000000, 0.000000, 0.000000}
\pgfsetfillcolor{diafillcolor}
\pgfsetfillopacity{1.000000}

\pgfsetarrowsend{latex}
\definecolor{dialinecolor}{rgb}{0.000000, 0.000000, 0.000000}
\pgfsetstrokecolor{dialinecolor}
\pgfsetstrokeopacity{1.000000}
\draw (7.000000\du,6.300000\du)--(7.000000\du,5.600000\du);
}
\pgfsetlinewidth{0.000000\du}
\pgfsetdash{}{0pt}
\pgfsetmiterjoin
{\pgfsetcornersarced{\pgfpoint{0.000000\du}{0.000000\du}}\definecolor{diafillcolor}{rgb}{0.000000, 0.000000, 0.000000}
\pgfsetfillcolor{diafillcolor}
\pgfsetfillopacity{0.000000}
\fill (1.611430\du,7.088820\du)--(1.611430\du,8.488820\du)--(6.202015\du,8.488820\du)--(6.202015\du,7.088820\du)--cycle;
}{\pgfsetcornersarced{\pgfpoint{0.000000\du}{0.000000\du}}\definecolor{dialinecolor}{rgb}{0.000000, 0.000000, 0.000000}
\pgfsetstrokecolor{dialinecolor}
\pgfsetstrokeopacity{0.000000}
\draw (1.611430\du,7.088820\du)--(1.611430\du,8.488820\du)--(6.202015\du,8.488820\du)--(6.202015\du,7.088820\du)--cycle;
}% setfont left to latex
\definecolor{dialinecolor}{rgb}{0.000000, 0.000000, 0.000000}
\pgfsetstrokecolor{dialinecolor}
\pgfsetstrokeopacity{1.000000}
\definecolor{diafillcolor}{rgb}{0.000000, 0.000000, 0.000000}
\pgfsetfillcolor{diafillcolor}
\pgfsetfillopacity{1.000000}
\node[anchor=base east,inner sep=0pt, outer sep=0pt,color=dialinecolor] at (6.002015\du,7.982882\du){$\vc_*^a$};
\pgfsetlinewidth{0.000000\du}
\pgfsetdash{}{0pt}
\pgfsetmiterjoin
{\pgfsetcornersarced{\pgfpoint{0.000000\du}{0.000000\du}}\definecolor{diafillcolor}{rgb}{0.000000, 0.000000, 0.000000}
\pgfsetfillcolor{diafillcolor}
\pgfsetfillopacity{0.000000}
\fill (5.400000\du,6.400000\du)--(5.400000\du,7.200000\du)--(8.592500\du,7.200000\du)--(8.592500\du,6.400000\du)--cycle;
}{\pgfsetcornersarced{\pgfpoint{0.000000\du}{0.000000\du}}\definecolor{dialinecolor}{rgb}{0.000000, 0.000000, 0.000000}
\pgfsetstrokecolor{dialinecolor}
\pgfsetstrokeopacity{0.000000}
\draw (5.400000\du,6.400000\du)--(5.400000\du,7.200000\du)--(8.592500\du,7.200000\du)--(8.592500\du,6.400000\du)--cycle;
}% setfont left to latex
\definecolor{dialinecolor}{rgb}{0.000000, 0.000000, 0.000000}
\pgfsetstrokecolor{dialinecolor}
\pgfsetstrokeopacity{1.000000}
\definecolor{diafillcolor}{rgb}{0.000000, 0.000000, 0.000000}
\pgfsetfillcolor{diafillcolor}
\pgfsetfillopacity{1.000000}
\node[anchor=base,inner sep=0pt, outer sep=0pt,color=dialinecolor] at (6.996250\du,6.994063\du){$\vc_*^l$};
\pgfsetlinewidth{0.100000\du}
\pgfsetdash{}{0pt}
\pgfsetmiterjoin
{\pgfsetcornersarced{\pgfpoint{0.000000\du}{0.000000\du}}\definecolor{diafillcolor}{rgb}{1.000000, 0.886275, 0.713726}
\pgfsetfillcolor{diafillcolor}
\pgfsetfillopacity{1.000000}
\fill (4.900000\du,8.300000\du)--(4.900000\du,9.200000\du)--(8.525688\du,9.200000\du)--(8.525688\du,8.300000\du)--cycle;
}{\pgfsetcornersarced{\pgfpoint{0.000000\du}{0.000000\du}}\definecolor{dialinecolor}{rgb}{0.000000, 0.000000, 0.000000}
\pgfsetstrokecolor{dialinecolor}
\pgfsetstrokeopacity{1.000000}
\draw (4.900000\du,8.300000\du)--(4.900000\du,9.200000\du)--(8.525688\du,9.200000\du)--(8.525688\du,8.300000\du)--cycle;
}% setfont left to latex
\definecolor{dialinecolor}{rgb}{0.000000, 0.000000, 0.000000}
\pgfsetstrokecolor{dialinecolor}
\pgfsetstrokeopacity{1.000000}
\definecolor{diafillcolor}{rgb}{0.000000, 0.000000, 0.000000}
\pgfsetfillcolor{diafillcolor}
\pgfsetfillopacity{1.000000}
\node[anchor=base,inner sep=0pt, outer sep=0pt,color=dialinecolor] at (6.712844\du,8.944063\du){attention};
\pgfsetlinewidth{0.100000\du}
\pgfsetdash{}{0pt}
\pgfsetmiterjoin
{\pgfsetcornersarced{\pgfpoint{0.000000\du}{0.000000\du}}\definecolor{diafillcolor}{rgb}{1.000000, 0.843137, 0.870588}
\pgfsetfillcolor{diafillcolor}
\pgfsetfillopacity{0.000000}
\fill (5.833830\du,4.173760\du)--(5.833830\du,5.726926\du)--(7.318673\du,5.726926\du)--(7.318673\du,4.173760\du)--cycle;
}{\pgfsetcornersarced{\pgfpoint{0.000000\du}{0.000000\du}}\definecolor{dialinecolor}{rgb}{0.000000, 0.000000, 0.000000}
\pgfsetstrokecolor{dialinecolor}
\pgfsetstrokeopacity{0.000000}
\draw (5.833830\du,4.173760\du)--(5.833830\du,5.726926\du)--(7.318673\du,5.726926\du)--(7.318673\du,4.173760\du)--cycle;
}% setfont left to latex
\definecolor{dialinecolor}{rgb}{0.000000, 0.000000, 0.000000}
\pgfsetstrokecolor{dialinecolor}
\pgfsetstrokeopacity{1.000000}
\definecolor{diafillcolor}{rgb}{0.000000, 0.000000, 0.000000}
\pgfsetfillcolor{diafillcolor}
\pgfsetfillopacity{1.000000}
\node[anchor=base,inner sep=0pt, outer sep=0pt,color=dialinecolor] at (6.576252\du,5.144406\du){OR};
\end{tikzpicture}
}
\end{center}
\caption[]{\label{fig:las_jatd} Details of JATD implementation. The ``JATD'' block can be added to the outputs of either attention block in fig. \ref{fig:delib}.
}
\end{figure}

LAS-JATD \cite{sainath2020jatd} augments LAS to enable training on paired audio-text data, as well as unpaired (i.e. text-only) data. This is done through the introduction of a new learnable fixed context vector, $\vc_e^l$, which is used as an alternative to the acoustic context vector $\vc_e^a$ (fig. \ref{fig:las_jatd}). During inference, two log probabilities are produced, one based on $\vc_e^a$ and the other based on $\vc_e^l$. These are interpolated using weight $\lambda$ to produce the final output log probabilities:

\begin{equation}
\label{eq:las_jatd_probs}
\lambda \log{p(\vy_u^d |\vx, \vc_e^a , \vy_{u - 1:1}^d )} + (1 - \lambda)\log{p(\vy_u^d | \vc_e^l , \vy_{u-1:1}^d )}
\end{equation}

The first term in this equation represents the familiar acoustic model (i.e. a regular two-pass LAS). The second term, $\log{p(\vy_u^d | \vc_e^l , \vy_{u-1:1}^d )}$, can be thought of as a language model since it does not depend on acoustic features, $\vx$.

LAS-JATD also provides a framework for incorporating unpaired data into training (eq. \ref{eq:las_jatd_loss}). It uses both acoustic and learnable context vectors when training on both paired and unpaired data. Acoustic context vectors are generated based on real audio, $\vx^a \in \vx$, for paired examples and ``created'' audio, $\vx^l \in \vx$, for unpaired examples. Importantly, it restricts training so only paired examples update the encoder attention parameters and only unpaired examples update the fixed context vector. This avoids biasing acoustic attention parameters towards unpaired data, and was found to be effective in \cite{sainath2020jatd}. In this work, we explore synthesizing audio based on text data to create $\vx^l$.

\begin{equation}
\label{eq:las_jatd_loss}
  \mathcal{L} =
    \begin{cases}
      \lambda \log{p(\vy_u^d |\vx^a, \vc_e^a , \vy_{u - 1:1}^d )} + (1 - \lambda)\log{p(\vy_u^d | \vc_e^l , \vy_{u-1:1}^d )}\\
      \text{if paired example}\\
      \lambda \log{p(\vy_u^d |\vx^l, \vc_e^a , \vy_{u - 1:1}^d )} + (1 - \lambda)\log{p(\vy_u^d | \vc_e^l , \vy_{u-1:1}^d )}\\
      \text{if unpaired example}\\
    \end{cases} 
\end{equation}

LAS-JATD improves performance through the addition of unpaired data to its training set, but misses out on gains from the spell-correcting capabilities of deliberation models.

\subsection{Proposed Method: Deliberation-JATD
\label{sec:methods_delib_jatd}}

We propose the Deliberation-JATD model, combining deliberation's spell-correcting benefits with JATD's ability to train on unpaired data. Similar to LAS-JATD, this model uses fixed context vectors as an alternative to attention context vectors, $c_e^a$ and $c_b^a$. This results in a new set of output log probabilities that act as a language model.

Given that deliberation models contain two attention contexts, we examine two Deliberation-JATD variations. The first, dubbed ``partial JATD'', uses the fixed context vector $\vc_e^l$ as an alternative to the encoder attention context, $\vc_e^a$, while continuing to use first-pass decoder context, $\vc_b^a$. This results in the LM log probabilities, $\log{p(\vy_u^d |\vx, \vc_e^l , \vc_b^a, \vy_{u - 1:1}^d )}$, and the following final model outputs used during inference:

\begin{equation}
\label{eq:deliberation_partial_jatd_probs}
\lambda \log{p(\vy_u^d |\vx, \vc_e^a , \vc_b^a, \vy_{u - 1:1}^d )} + (1 - \lambda)\log{p(\vy_u^d |\vx, \vc_e^l , \vc_b^a, \vy_{u - 1:1}^d )}
\end{equation}

The second variant, named ``full JATD'', goes one step further and adds a second fixed context vector, $\vc_b^l$, as an alternative to its first pass decoder attention context, $\vc_b^a$ (eq. \ref{eq:deliberation_full_jatd_probs}). This means that both attention contexts are replaced and the LM log probabilities having no dependence on the acoustic inputs, $\vx$.

\begin{equation}
\label{eq:deliberation_full_jatd_probs}
\lambda \log{p(\vy_u^d |\vx, \vc_e^a , \vc_b^a, \vy_{u - 1:1}^d )} + (1 - \lambda)\log{p(\vy_u^d | \vc_e^l , \vc_b^l, \vy_{u - 1:1}^d )}
\end{equation}

We use the ``joint'' training strategy \cite{sainath2020jatd} to train both variants on a mix of supervised audio and text-only data paired with TTS audio. The resulting training loss is similar to eq. \ref{eq:las_jatd_loss}, but incorporates eq. \ref{eq:deliberation_partial_jatd_probs} or eq. \ref{eq:deliberation_full_jatd_probs} for the unpaired text term, depending on the model variation used. Similar to LAS-JATD, we distinguish between real audio from paired data, $\vx^a$, and synthetic audio from text data, $\vx^l$, and only backpropagate some parameters for each type of data. Specifically, encoder attention parameters are held constant for synthetic audio and the fixed attention context vectors are held constant for real audio. The second-pass decoder parameters are updated for both types of data.

Full JATD's LM is similar to LAS-JATD's LM in that it replaces all attention contexts (from the encoder and first-pass decoder) with fixed vectors. This means that the LM component's second-pass decoder completely ignores the first-pass decoder and all acoustic information. As a result, the LM learns to spell-correct based purely on previous model outputs, $\vy_{u - 1:1}^d$. In contrast, partial JATD was designed to more frequently expose the second-pass decoder to the first-pass decoder during training. Its LM keeps the first-pass decoder attention context, allowing the second-pass decoder to train with it and the LM to make use of some indirect acoustic information.

\subsection{Training
\label{sec:methods_training}}

All models were initialized from an RNN-T trained with supervised audio-text paired data. The same model was also used as the baseline RNN-T model when analyzing performance. All training sets that included TTS used a mix of 90\% audio-text pairs and 10\% pure text data with corresponding TTS. We will refer to this training set as the ``mixed audio'' training set. For all JATD models, an interpolation weight of $\lambda = 0.1$ was found to work well in training.

\section{Experiment Details \label{sec:experiments}}

\subsection{Training Sets}

We use the same paired audio-text training set as \cite{Arun19}. This dataset consists of approximately 180M multi-domain utterances spanning domains of search, farfield, telephony and YouTube English utterances. The search and farfield utterances are anonymized and hand-transcribed and are representative of Google’s voice search traffic.

Our unpaired data consisted of 4.6M samples of query text from anonymized Google Maps traffic. We paired this text with synthesized audio generated by a multi-speaker TTS system based on the architecture described in \cite{peyser2019numeric}. Our synthesized audio is generated in the voice of 98 English speakers covering American, Australian, British and Singaporean accents with a Google Assistant clean speaking style. Each utterance's audio was synthesized using a single voice assigned randomly during training set generation. 

Both real and synthesized audio training data were artificially corrupted using a room simulator. Various degrees of noise and reverberation were added such that the overall SNR is between 0dB and 30dB, with an average SNR of 12dB \cite{Chanwoo17}. The noise sources were from YouTube and noisy environmental recordings.

\subsection{Test Sets}
We use two primary evaluation sets: ``VS'' and ``SXS''. VS consists of ~14k anonymized hand-transcribed utterances from Google traffic. SXS is a sample of 1,200 real-audio utterances where the conventional model \cite{Golan16} outperformed the E2E LAS rescoring model \cite{SainathPang19}. This dataset consists largely of rare words and is useful for measuring how including text-only training data can improve some of these errors.

We used a corpus composed purely of rare word utterances \cite{peyser2020fusion} to specifically evaluate model performance on rare nouns. This corpus is a non-overlapping random subset of the same text data that was sampled to create the unpaired training data. It consists of utterances containing words that (1) occurred either once or not at all in the paired training set and (2) accounted for less than 1 in every million words in the text data. For privacy reasons, we only included words that occurred more than 1000 times in the text data.

We created two evaluation sets derived from this rare word corpus. The ``TTS'' set is a subset of 10,000 of these utterances combined with synthetic audio generated by the same system that synthesized audio for the training set in \cite{peyser2019numeric}. For this test set, the audio synthesizing system was configured to use a voice profile distinct from all those used in the training set. The ``Spoken Text'' evaluation set is a random sample of 200 utterances from the TTS set, but manually spoken and recorded by one of our team members. This set was used to verify that improvements in model performance on the TTS evaluation sets were matched on real audio.

\subsection{Architecture Details 
\label{sec:experiments_modeling}}
We used the same deliberation model configuration as \cite{hu2020deliberation}. All experiments used 128-dimensional log-Mel features, computed with a 32ms window and shifted every 10ms. Similar to \cite{Arun19}, features for each frame are stacked with 3 frames to the left and then downsampled by 3 to a 30ms frame rate.

The first-pass RNN-T network is similar to \cite{SainathPang19}, consisting of an 8 LSTM layer encoder and 2 LSTM layer prediction network. Each LSTM layer has 2,048 hidden units followed by a 640-dimensional projection layer. There is a factor of 2 time-reduction layer after the second encoder LSTM layer. The outputs of encoder and prediction network are fed to a 640 hidden unit joint network followed by a softmax layer predicting 4,096 lowercase wordpieces.

The first-pass RNN-T hypotheses are padded with end-of-sentence label $\langle \backslash \text{s} \rangle$ to a length of 120. Each subword in a hypothesis is mapped to a vector by a 96-dimensional embedding layer and encoded by a 2-layer bidirectional LSTM encoder, where each layer has 2,048 hidden units followed by a 320-dimensional projection. Both attention models use multi-headed attention \cite{Vaswani17} with four attention heads. The two output context vectors are concatenated and fed to a 2-layer LSTM decoder (2,048 hidden units followed by a 640-dimensional projection per layer). The second-pass attention decoder has a 4,096-dimensional softmax layer to predict the same mixed-case wordpieces \cite{schuster2012japanese} as the RNN-T. We use a similar architecture for our LAS models. The second-pass decoder in these models is the same as before (2 LSTM layers, each with 2,048 hidden units followed by a 640-dimensional projection).

The total size of the RNN-T model is 114M parameters, and the second-pass decoder has 33M parameters. All models are trained in Tensorflow \cite{AbadiAgarwalBarhamEtAl15} using the Lingvo \cite{shen2019lingvo} toolkit on a v2-128 Cloud TPU slice with a global batch size of 4,096.

For JATD models, the interpolation weight, $\lambda$, for inference is chosen (between 0.01, 0.025, and 0.05) to optimize WER for the VS test set. Results shown in table \ref{table:results_main} use $\lambda = 0.025$ for Deliberation-JATD and $\lambda = 0.01$ for LAS-JATD.

\section{Results \label{sec:results}}

\begin{table*} [t]
  \centering
  \begin{tabular}{|l||l|c|c|c|c|c|} \hline
    \multirow{2}{*}{ID} & \multirow{2}{*}{Model} & \multirow{2}{*}{Training Data} & \multicolumn{4}{c|}{WER (\%)}  \\
    \cline{4-7}
    & & & VS & SXS & TTS & Spoken Text \\ \hline
    \textit{B0} & RNN-T & Paired & 6.6 & 28.3 & 41.9 & 26.9 \\ \hline
    \textit{B1} & RNN-T & Mixed & 6.2 & 30.4 & 39.6 & 22.5 \\ \hline
    \textit{B2} & LAS & Paired & 6.7 & 25.6 & 41.4 & 27.4 \\ \hline
    \textit{B3} & LAS & Mixed & 6.7 & 24.6 & 38.4 & 24.1 \\ \hline
    \textit{B4} & LAS-JATD & Mixed & 7.0 & 26.1 & \underline{33.1} & 24.0 \\ \hline
    \textit{B5} & Deliberation & Paired & \underline{\textbf{5.7}} & 22.0 & 39.5 & 22.2 \\ \hline
    \textit{B6} & Deliberation & Mixed & 5.8 & \underline{21.9} & 34.8 & \underline{20.5} \\ \hline
    \hline
    \textit{E0} & Deliberation-JATD (Partial) & Mixed & 5.8 & 21.9 & 30.9 & 20.7 \\ \hline
    \textit{E1} & Deliberation-JATD (Full) & Mixed & \textbf{5.7} & \textbf{21.7} & \textbf{30.6} & \textbf{18.7} \\ \hline
  \end{tabular}
  \caption{Model performance comparison. Lowest baseline WER values are \underline{underlined}, lowest overall WER values are \textbf{bolded}.
}
  \label{table:results_main}
  \vspace{-0.1in}
\end{table*}

We now analyze the performance of our Deliberation-JATD model. Table \ref{table:results_main} compares the performance of our two model variants (E0, E1) with a set of baseline models (B0 to B6). All models are trained on either paired data or the mixed audio training set (described in section \ref{sec:methods_training}).

B0 is an RNN-T model and serves as a baseline trained on paired data. It is also the model we used to initialize all LAS and deliberation variants. B1 has the same architecture as B0, but trained from scratch on the mixed audio training set. The addition of TTS data to training improves RNN-T performance on the TTS, VS, and Spoken Text test sets while degrading on the SXS set.

B2 is an LAS model trained on paired data and B3 is the same model trained on the mixed audio training set. B4 is our LAS-JATD implementation. Training on mixed audio (B3) provides some modest performance improvements compared to B2. The LAS-JATD model shows the lowest WER among baseline models on the TTS set. It also improves on the Spoken Text relative to regular LAS, but degrades VS and SXS performance.

B5 and B6 are implementations of the two-pass deliberation model trained on the paired and mixed audio training sets, respectively. They show the strongest metrics on Spoken Text as well as the VS and SXS sets. They also have the lowest TTS WER among non-JATD models. Training deliberation on the mixed audio (B6), as opposed to only the paired data (B5), results in significant improvements on all but the VS test set.

We compare the aforementioned baselines against our Deliberation-JATD models: the full variant (E0) and the partial variant (E1), both trained on the mixed audio training set. Both variants show significant gains on all sets sets aside from VS, which is roughly unchanged. The full variant (E1) produces the lowest WER of all models on the SXS, TTS, and Spoken Text test sets and matches the lowest WER obtained on VS by B5. On the TTS test set, it shows a 22.5\% improvement relative to deliberation trained on paired data (B5), and a 12\% improvement relative to deliberation trained on mixed data (B6). Similar gains are seen on the Spoken Text set.

Comparing the Deliberation-JATD models, we notice that the full variant outperforms the partial variant despite the fact that the full JATD LM term (described in section \ref{sec:methods_delib_jatd}) ignores the first-pass decoder outputs, while the partial JATD LM term uses them. We speculate that partial JATD would benefit from a real/TTS bit passed to the bidirectional LSTMs that encode the first-pass decoder output. This would allow its LM component to distinguish between paired and TTS audio. We leave this as future work.

\definecolor{dark_red}{rgb}{0.8, 0, 0}
\definecolor{dark_green}{rgb}{0, 0.55, 0}
\newcommand{\ERR}[1]{{\color{dark_red}\textbf{#1}}}
\newcommand{\CORR}[1]{{\color{dark_green}\textbf{#1}}}

\begin{table} [h!]
  \centering
  \begin{tabular}{ll} \hline
    Deliberation (B5) & Full Deliberation-JATD (E1) \\ \hline \hline
    \ERR{tough trees} leasing office & \CORR{toftrees} leasing office \\ \hline
    \ERR{chow mein} jackson & \CORR{cal-maine} jackson \\
    mississippi & mississippi \\ \hline
    distance from \ERR{wanderleo} & distance from \CORR{juan dolio} \\
    to punta cana & to punta cana \\ \hline
    \ERR{nellis ford} realty & \CORR{nellysford} realty \\ \hline \hline
    \CORR{southline} & \ERR{south lyon} \\ \hline
    the mansions of & the mansions of \\
    \CORR{shadowbriar} & \ERR{shadow briar} \\
    houston texas & houston texas \\ \hline
    \CORR{delias} near me & \ERR{delia's} near me \\ \hline
  \end{tabular}
  \caption{Sample wins and losses comparing full deliberation-JATD (E1) and deliberation (B5) on the Spoken Text test set. Correct and incorrect portions highlighted in green and red, respectively.}
  \label{table:results_err}
  \vspace{-0.1in}
\end{table}

Finally, Table \ref{table:results_err} shows a sample of wins and losses when comparing deliberation (B5) to the the full variant of deliberation-JATD (E2). The deliberation-JATD wins mostly by correcting transcription errors for proper nouns such as ``toftrees'' and ``nellysford''. The losses are sometimes also related to proper nouns (e.g. ``southline'' to ``south lyon''), but mostly due to spelling errors, e.g. ``delia's'' in place of ``delias''.

\section{Conclusions \label{sec:conclusions}}

We presented a new Deliberation-JATD model, which incorporates unpaired text data in a deliberation model training jointly with acoustic (i.e. paired) data. The proposed method significantly outperforms both Deliberation and LAS-JATD models, reducing WER by up to 22.5\% relative to a regular deliberation model \cite{hu2020deliberation} on a rare word test set. Although the regular deliberation is improved by training from scratch using both paired and unpaired data, it still lags behind the Deliberation-JATD model by 12\% in terms of WER. The superior performance of Deliberation-JATD is achieved without additional inference complexity, mult-stage training, or performance degradation on Google Voice Search tasks.

\section{Acknowledgements}
Thank you to Ruoming Pang and Cal Peyser with their help in producing training and test sets used for this work.
\newpage

\bibliographystyle{IEEEbib}
\bibliography{main}
\end{document}